\documentclass{jpsj-suppl}
\usepackage[utf8]{inputenc}
\usepackage{txfonts} 
\usepackage{xspace}

\def\ppbar{\mbox{(anti-)}proton\xspace}
\def\ppbars{\mbox{(anti-)}protons\xspace}

\title{Towards Quantum Logic Inspired Cooling and Detection for Single (Anti-)Protons}

\author{Teresa \textsc{Meiners}$^{1,2}$, Malte \textsc{Niemann}$^{1,2}$, Anna-Greta \textsc{Paschke}$^{1,3}$, Johannes \textsc{Mielke}$^{1}$, Alexander \textsc{Idel}$^{1}$, Matthias \textsc{Borchert}$^{1}$, Kai \textsc{Voges}$^{1}$, Amado \textsc{Bautista-Salvador}$^{1,2,3}$, Stefan \textsc{Ulmer}$^{4}$, Christian \textsc{Ospelkaus}$^{1,2,3}$}

\inst{
$^{1}$Institut f\"ur Quantenoptik, Leibniz Universit\"at Hannover,\\
Welfengarten 1, 30167 Hannover, Germany\\
$^{2}$Laboratory of Nano and Quantum Engineering, Leibniz Universit\"at Hannover,\\
Schneiderberg 39, 30167 Hannover\\
$^{3}$Physikalisch-Technische Bundesanstalt, Bundesallee 100, 38116 Braunschweig, Germany\\
$^{4}$Ulmer Initiative Research Unit, RIKEN, 2-1 Hirosawa, Wako, Saitama 351-0198, Japan 
}

\email{christian.ospelkaus@iqo.uni-hannover.de}

\recdate{May 31, 2016}

\abst{We discuss laser-based and quantum logic inspired cooling and detection methods amenable to single \ppbars. These would be applicable e.~g.~in a g-factor based test of CPT invariance as currently pursued within the BASE collaboration. Towards this end, we explore sympathetic cooling of single \ppbars with atomic ions as suggested by Heinzen and Wineland (1990). }

\kword{laser cooling, Penning traps, atomic ions, g-factor, proton, antiproton}

\begin{document}
\maketitle

\section{Introduction}

Penning trap precision measurements of masses or $g$-factors employ either destructive or non-destructive induced image charge (IIC) detection of particles. In most experiments, cooling of the particles and thus control over their motion is performed through the interaction with a cryogenic tank circuit. A challenging example of Penning trap precision measurements are single protons and antiprotons. Comparisons of the $g$-factor and $q/m$ of single protons and antiprotons for testing the fundamental CPT symmetry have progressed rapidly~\cite{ulmer_observation_2011,mooser_resolution_2013,disciacca_direct_2012,mooser_demonstration_2013,atrap_collaboration_one-particle_2013,mooser_direct_2014,ulmer_high-precision_2015}. These experiments employ IIC and the continuous Stern-Gerlach effect~\cite{dehmelt_proposed_1973} for spin state detection and suffer from the difficulty to cool the particles efficiently below 4\,K~\cite{mooser_resolution_2013}. Heinzen and Wineland~\cite{heinzen_quantum-limited_1990} proposed a method set, which would allow coupling of in principle arbitrary charged particles of interest to a laser-cooled atomic ion for sympathetic laser cooling and state readout. Here, we describe progress towards sympathetic cooling and quantum logic detection of single \ppbars by a laser-cooled $^9$Be$^+$ ion. 

\section{Trap setup}
\begin{figure}[htb]
	\begin{minipage}{\columnwidth}
		\centering
		\includegraphics[width=0.9\columnwidth]{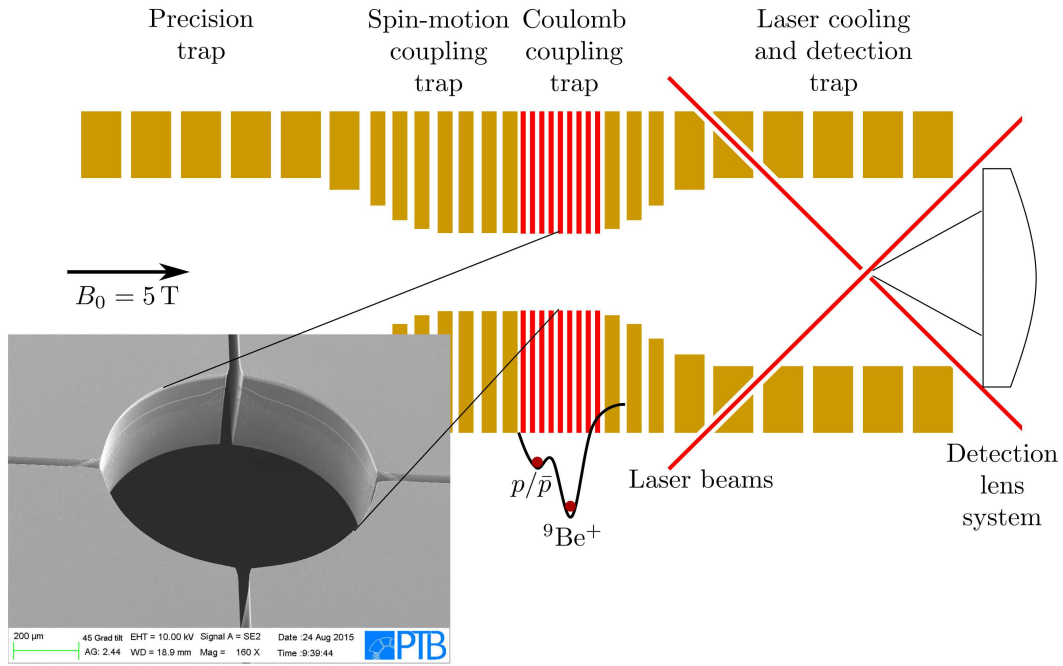}
	\end{minipage}
	\caption{Illustration of trap stack (see main text). The scanning electron microscope image shows a sample structure for the Coulomb coupling trap. The structure is etched out of a $380\,\mu\mathrm{m}$ thick silicon wafer using deep reactive ion etching (DRIE). After subsequent metallization, the side walls of the inner $800\,\mu\mathrm{m}$ diameter hole will form the electrodes of a cylindrical Penning trap. Multiple of similar structures, stacked on top of each other with an insulating spacer in between, will produce a double-well potential to couple a single \ppbar to a single $^9$Be$^+$ ion. Segmented electrodes such as the ones depicted here can be used to implement e.~g.~an axialization drive. Note that the wafer for the final trap stack will have a thickness of $200\,\mu\mathrm{m}$.}
	\label{fig:trap_stack_setup}
\end{figure}

Experiments will be carried out in a cylindrically symmetric Penning trap array (magnetic field $B_0=5\,\mathrm{T}$). Parts of the trap array are based on the BASE CERN apparatus~\cite{smorra_base_2015}. The trap stack will consist of modules with dedicated functions: 
\begin{itemize}
	\item A {\it precision trap} where motional and spin resonances of single \ppbars will be probed using resonant excitations.
	\item A {\it spin-motion coupling trap} where spin and motional degrees of single \ppbars can be coupled in order to detect the spin state of the \ppbar.
	\item A {\it Coulomb coupling trap} consisting of a double-well potential for a single \ppbar and a single $^9$Be$^+$ ion, allowing swapping of motional states between the two particles for cooling and state readout.
	\item A {\it laser cooling and detection trap} for manipulation and readout of a single $^9$Be$^+$ ion.
\end{itemize}
Particles will be shuttled between these trap modules to perform different operations. The trap stack is illustrated in Fig.~\ref{fig:trap_stack_setup}.

We have designed a {\it laser cooling and detection trap} with 9\,mm inner diameter made from gold plated OFHC copper with sapphire spacers. It features 4 laser ports at 45 degrees with respect to the trap axis, so that beams for laser cooling and detection can be delivered to the ion. Detection will be realized via a lens system downstream from the trap and collinear with the trap axis. The fluorescence light emitted from the ion will be imaged to a focus outside the vacuum system and detected on a CCD camera for alignment or on a PMT for quantitative data taking. In order to load single ions into the trap, we will employ a scheme first demonstrated with calcium ions~\cite{hendricks_all-optical_2007} and recently adapted in our laboratory for $^9$Be$^+$ ions in surface-electrode traps in the context of trapped-ion quantum information processing~\cite{wahnschaffe_exploring_2016}. We shine a  single nanosecond 1064\,nm laser pulse onto a solid Beryllium target to release a gas jet of neutral $^9$Be atoms. Light at 235\,nm is used to ionize the atoms in a two-photon process connecting the $^1S$ to the $^1P$ state and then to the continuum (see Fig.~\ref{fig:pi_laser}a)). The advantage of this ``push of a button ion loading'' approach is that it will introduce minimal heating in the cryogenic environment and minimize the time that the deep UV 235\,nm laser beam is on. In our experiments with surface-electrode traps, we have seen that this minimizes the detrimental charging of dielectrics as a result of ion loading. 

The {\it Coulomb coupling trap} will consist of a stack of micro-structured and metallized wafers with a thickness of $\approx200\,\mu\mathrm{m}$ and an inner diameter of $\approx800\,\mu\mathrm{m}$. Thicker test samples have been fabricated using deep reactive ion etching of silicon~\cite{maluf_introduction_2004}. The scanning electron microscope picture in Fig.~\ref{fig:trap_stack_setup} shows a structure where the inner hole of a cylindrical micro Penning trap has been etched, including segmented side walls, which might be desirable to apply different motional-mode couplings. A stack of 9 similar electrodes will suffice to generate a double-well potential with a trap frequency of $2\pi\cdot4\,\mathrm{MHz}$ per well for each of the \ppbar and the $^9$Be$^+$ ion at a separation of $300\,\mu\mathrm{m}$ between them. This allows the two particles to fully swap their motional states within 3.7\,ms for state detection and sympathetic cooling. This mechanism has been demonstrated for pairs of atomic ions at the level of single quanta of motion in surface-electrode Paul traps~\cite{brown_coupled_2011,harlander_trapped-ion_2011}. 

The {\it spin-motional coupling trap} will provide a mechanism to map the spin state of a single \ppbar into its motional state in order to later transfer it to the atomic ion in the Coulomb coupling trap for subsequent spin state detection. As the wavelength of the spin-flip transition for the \ppbar is rather long compared to the spatial extent of its motion, special measures need to be taken in order to be able to jointly (de-)excite motional and internal degrees of freedom of the particle. Similar protocols have been implemented with atomic ions in the context of quantum logic operations using microwave radiation \cite{ospelkaus_microwave_2011,khromova_designer_2012} and have been adapted for the cylindrical Penning trap geometry of the present project. 

\section{Laser systems}
\begin{figure}[htb]
	\includegraphics[width=\columnwidth]{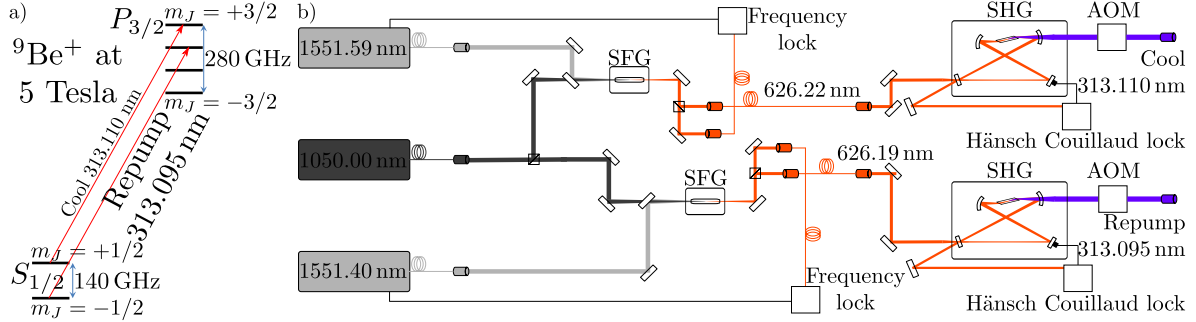}
	\caption{a) Energy structure of $^9$Be$^+$ at 5\,Tesla. The distance between the $S_{1/2}$ and $P_{3/2}$ energy levels is not to scale. b) Doppler cooling laser system for $^9$Be$^+$ ions. Two ultraviolet laser beams near 313\,nm are used for Doppler cooling and repumping. }
	\label{fig:doppler_cooling_laser}
\end{figure}

Laser systems for cooling, state detection and ionization of $^9$Be$^+$ have been set up. Laser cooling will be performed on the $\left|S_{1/2}, m_J=+1/2,m_I=+3/2\right>$ $\rightarrow$ $\left|P_{3/2}, m_J=+3/2,m_I=+3/2\right>$ transition (``cycling transition''). Population trapped in the $\left|S_{1/2}, m_J=-1/2,m_I=+3/2\right>$ state will be pumped back into the detection cycle using a laser beam resonant with the $\left|S_{1/2}, m_J=-1/2,m_I=+3/2\right>$ $\rightarrow$ $\left|P_{3/2}, m_J=+1/2,m_I=+3/2\right>$ (``repump'') transition. The energy levels and transitions are shown in Fig.~\ref{fig:doppler_cooling_laser}a). To generate the two separate laser beams near 313\,nm, a laser system has been set up based on the scheme presented in~\cite{wilson_750-mW_2011}. A schematic of the laser system is shown in Fig.~\ref{fig:doppler_cooling_laser}b). A single laser beam at 1050\,nm is split into two beam paths. Each of these beams is superimposed with a laser beam from another laser near 1550\,nm to generate light near 626\,nm in a periodically poled lithium niobate (PPLN) nonlinear crystal via sum frequency generation (SFG). The resulting light is then frequency doubled to 313\,nm using second harmonic generation (SHG) in BBO crystals in a bow-tie enhancement cavity.

\begin{figure}[htb]
	\begin{minipage}{\columnwidth}
	\centering
	\includegraphics[width=0.6\columnwidth]{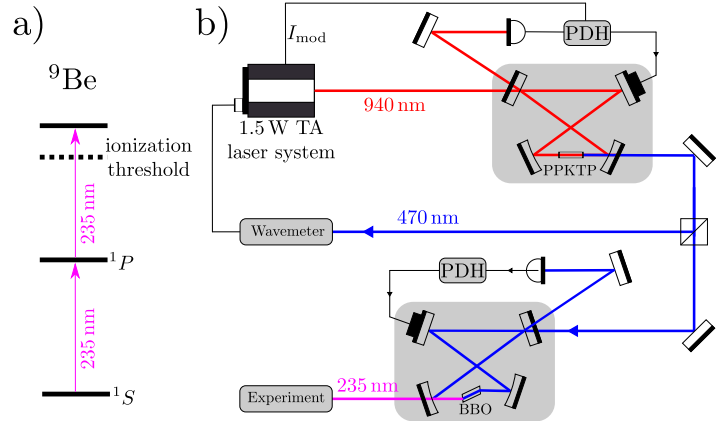}
	\end{minipage}
	\caption{Schematic of photoionization laser system. Light at 235\,nm resonant with the $^9$Be $^1S$ $\rightarrow$ $^1P$ transition is generated by frequency quadrupling a 940\,nm tapered amplifier laser source.}
	\label{fig:pi_laser}
\end{figure}

The photoionization light at 235\,nm is generated through frequency quadrupling of a 1.5\,W 940\,nm tapered amplifier laser source (cf.~\cite{lo_all-solid-state_2013}). The setup is shown in Fig.~\ref{fig:pi_laser}b). A first step of frequency doubling to 470\,nm has been shown to generate stable output power levels of up to 500\,mW. The second doubling stage from 470\,nm to 235\,nm generates power levels exceeding 20\,mW, whereas typically of order 1\,mW would be used for ionization in our surface-trap experiments. 

For ground state cooling in the Penning trap, the large ground state energy splitting of $\approx140\,\mathrm{GHz}$ between the two $m_J=\pm1/2$ sublevels of the $S_{1/2}$ state at 5\,T must be taken into account. Motional-sideband ground state cooling requires this frequency difference to be bridged using stimulated Raman transitions. However, the frequency difference between these two states is too large to be bridged using the established acousto-optic modulator technique. We are currently exploring an approach based on a mode-locked laser~\cite{campbell_ultrafast_2010,hayes_entanglement_2010}, where pairs of comb teeth in the frequency domain team up to provide stimulated Raman transitions. The system is based on a pulsed version of~\cite{wilson_750-mW_2011}. In the frequency doubling step from 626\,nm to 313\,nm, spectral compression~\cite{marangoni_narrow-bandwidth_2007} allows an efficient use of available power while maintaining a spectrally narrow pulse train with a width chosen to allow for the smallest detuning without accidentally inducing excessive spontaneous emission.

\section{Summary and outlook}
We have described steps towards sympathetic cooling and detection of single \ppbars using $^9$Be$^+$ ions. We are currently setting up an experiment to demonstrate this approach using protons. Since the design is adapted from the BASE experiment at CERN~\cite{smorra_base_2015}, it will be possible to implement these ideas at the BASE CERN setup with antiprotons once successfully demonstrated with protons. Our immediate focus is on implementing control over $^9$Be$^+$ in a cryogenic Penning trap. Further steps will comprise sympathetic Doppler cooling of single protons, ground state cooling and quantum logic spectroscopy of spin and motional resonances. 

\section{Acknowledgements}
We acknowledge helpful discussions with M.~Marangoni and G.~Cerullo, R.~Lehnert, and members of the BASE collaboration and the NIST ion storage group. We acknowledge financial support from ERC StG ``QLEDS'', QUEST and Leibniz Universit\"at Hannover. We are grateful for support by the PTB clean room facility team.

\end{document}